\documentclass{aa}
\usepackage{graphicx,times}

\begin{document}

\title{Detection of X-ray line emission from the 
shell of SNR B0540--69.3 with XMM-Newton RGS 
\thanks{Based on observations
obtained with XMM-Newton, an ESA science mission with instruments
and contributions directly funded by ESA Member States and the
USA (NASA).}
}

\author{K.\,J. van der Heyden \inst{1}
         \and
         F. Paerels \inst{2}
         \and
         J. Cottam \inst{2}
	 \and
         J. S. Kaastra \inst{1}
         \and
         G. Branduardi-Raymont \inst{3}
         }

\offprints{K.\,J. van der Heyden}
         
\institute{SRON Laboratory for Space Research,
              Sorbonnelaan 2, 3584 CA Utrecht, the Nether\-lands
           \and
           Columbia Astrophysics Laboratory, Columbia University, 538 West      
              120th Street, New York, NY 10027, USA
           \and
           Mullard Space Science Laboratory, University College London,
              Holmbury St. Mary, Dorking, Surrey RH5 6NT, UK
           }
           
\authorrunning{K.\,J. van der Heyden et al.}
\titlerunning{X-ray line emission from SNR 0540--69.3}

\date{Received 2 October 2000/ Accepted }

\abstract{
We present X-ray observations of PSR 0540--69.3 with the XMM-Newton 
observatory. The spectra obtained with the Reflection Grating 
Spectrometer reveal, 
for the first time, emission from ionized species of O, Ne and Fe 
originating from the SNR shell. Analysis of the emission line spectrum
allows us to derive estimates of the
temperature, ionization timescale, abundances, location, and velocity
of the emitting gas.
\keywords{ISM: individual: SNR 0540--69.3 --
ISM: supernova -- X-rays: ISM }
}

\maketitle

\section{Introduction}

SNR 0540--69.3 is one of only a few Crab-like supernova remnants (SNRs) with a shell.  
SNR0540--69.3 harbors a pulsar, PSR B0540--69.3, discovered in the soft X-ray band 
by Seward, Harnden, \& Helfand (\cite{seward1}). Like the Crab, it shows plerion emission 
from a synchrotron nebula powered by the embedded young 
pulsar (Chanan, Helfand \& Reynolds \cite{chanan}). From the period (50 ms) and spin down rate 
($4.79 \times 10^{-13} {\rm s\, s^{-1}}$) a   
a characteristic age of 1660 yr and rotational energy loss rate of 
$1.5{\times}10^{38}$ergs s$^{-1}$ have been derived (Seward, Harnden, \& Helfand \cite{seward1}).

Early optical observations (Mathewson et al. \cite{mathewson}) classified SNR 
0540--69.3 as a young oxygen-rich SNR with an 8\arcsec\ diameter shell, 
bright in 
[\ion{O}{iii}]. Mathewson et al. (\cite{mathewson}) also found a diffuse patch 
of [\ion{O}{iii}] $\sim$30\arcsec\ to the west of the ring which indicated that the 
remnant 
was larger than 8\arcsec. A {\it ROSAT} High Resolution Imager observation by Seward 
\& 
Harnden (1994) also revealed emission from well outside the central region, 
which they interpreted as from a patchy outer shell with a diameter of 55\arcsec. 
This 
shell contributes $\sim$20\% to the measured flux in the {\it ROSAT} 0.1--2 keV band. 
More recently Gotthelf \& Wang (\cite{gotthelf}) presented a high resolution {\it Chandra} HRC 
observation of PSR B0540--69.3 which clearly shows emission from the outer 
shell.  

In this letter, we report the results of observations of SNR 0540--69  
acquired with the Reflection Grating Spectrometer on board the XMM-Newton 
Observatory (den Herder et al. \cite{denherder}). These observations allow us for the first time to 
detect and identify emission lines from the remnant shell. Our spectral analysis 
allows us to derive values for the temperature, ionization timescale, 
abundances, location, and velocity of the emitting gas. 

\section{Observations \& Data Reduction}
PSR B0540 was observed repeatedly as part of the calibration of the Reflection Grating 
Spectrometer (RGS) during the Calibration/Performance Verification (Cal/PV) phase.
For the purpose of this letter, we have analyzed the 
126~ks data obtained on 2000 May 26, which is the longest observation with the lowest background.  
The observation was  
performed with the telescope rolled such 
that the RGS dispersion axis was aligned at 21.29\degr, clockwise on the sky, 
from celestial north. 
The data were initially processed with the XMM-Newton Science Analysis 
Software (SAS).  The spectra were extracted by applying spatial 
filters to the CCD image while a CCD pulse-height filter was applied to select 
the $m=-1$ spectral order. The remnant is located on the edge of a broad region 
of diffuse emission (Wang \& Helfand \cite{wang}) which complicates 
background determination.  However, the image is small enough on the detector that spatial 
regions offset from the source position can be used for background 
subtraction.  The background spectra were extracted by applying the same pulse height filters as used  
for the source, to spatial regions at either edge of the 
camera.

A response matrix appropriate to the spatial extent of the source was generated as follows.  
A spatial mask corresponding to the RGS aperture was imposed on the {\it Chandra} HRC image, 
and the intensity distribution was integrated over the RGS cross-dispersion direction.  
The resulting profile was convolved with the RGS point source response matrix, generated with the SAS task RGSRMFGEN.
The spectral analysis was performed using the SRON SPEX (Kaastra et al. \cite{kaastra}) package, which contains the 
MEKAL code (Mewe et al. \cite{mewe}) for modeling thermal emission.

\begin{figure}
\resizebox{\hsize}{!}{\includegraphics{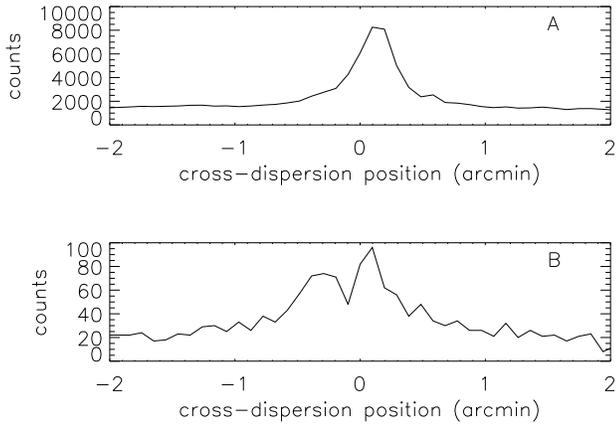}}
\caption[]{Cross-dispersion profiles of SNR 0540--69 obtained by integrating 
along the dispersion axis for a: the entire spectrum and b: \ion{O}{viii} 
Ly$\alpha$ only.}
\label{fig:fig1}
\end{figure}

\section{Analysis and Results}

\subsection{Location of emitting region}

We initially extracted spectra with spatial filters $0.25\arcmin$, 
$0.5\arcmin$, $1\arcmin$ and $1.5\arcmin$ wide, each centered about the peak emission. The 
most interesting feature in the spectra is the presence of an \ion{O}{viii} Ly$\alpha$
emission line that becomes more prominent as the width of the spatial extraction region is
increased.
This is an indication that at least 
some of the O emission is from the SNR shell, and not the central 
region. To further constrain the spatial distribution of the  emitting gas  we 
created a cross-dispersion image profile of the \ion{O}{viii} Ly$\alpha$
line by integrating over 
the line in the dispersion direction. The cross-dispersion profile, 
plotted in Fig.~\ref{fig:fig1}, shows two peaks.
Referring to the {\it Chandra} HRC image
(Gotthelf \& Wang \cite{gotthelf}, reproduced here as Fig. 2), we conclude that we see
O emission both from the central region of the remnant, centered on the pulsar, 
(the peak at offset zero in Fig. 1b), as well as from the partial shell to the west
(the second peak $\approx$ 24 arcsec towards more negative offset in Fig. 1b).

The angular extent of this partial shell, $\Delta\phi \approx 30\arcsec$, corresponds to 
an effective wavelength width along the dispersion direction of $\approx 0.06$ \AA, 
comparable to the wavelength resolution of the spectrometer. This implies that we
do not have the sensitivity to resolve variations in brightness along the shell.

\subsection{Spectral analysis}
We divided the remnant into three regions, Northeast (NE), Central (C) and 
Southwest (SW), for the purpose of spectral analysis. For the NE and SW regions we use 
an extraction width of 0.5\arcmin centered at $\pm 0.375 \arcmin$
about the central 
compact region, while for the brighter C region we use a width of 
0.25\arcmin. The spectra for these regions are displayed in Figs. 3 and 4.
Each  
spectrum shows a strong power law continuum with superimposed line emission, 
most notably \ion{O}{viii} Ly$\alpha$. 
Emission line features are weakest in the spectrum extracted from the NE 
region, where we only detect \ion{O}{viii} Ly$\alpha$ and possibly weak 
\ion{O}{vii} emission.
The line emission is strongest in the SW spectrum where we easily
identify emission lines of \ion{O}{viii} (Ly$\alpha$, Ly$\beta$), 
\ion{O}{vii} ($n=2-1$ triplet, He$\beta$), 
\ion{Fe}{xvii} ($15.01, 16.78, 17.05, 17.10$ \AA),
\ion{Ne}{x} (Ly$\alpha$), and \ion{Ne}{ix} ($n=2-1$ triplet). Due to 
galactic absorption, no 
features can be seen 
longward of $\sim 23$ \AA. Note that even though the shell-like
structure and the plerion are only partially resolved by the
telescopes of {\it XMM-Newton}, the high spectral resolution of the
RGS allows for high sensitivity to the discrete emission from the
remnant, even though the spectrum is dominated by the powerlaw
continuum from the plerion.

\begin{figure}
\resizebox{\hsize}{!}{\includegraphics[angle=-90]{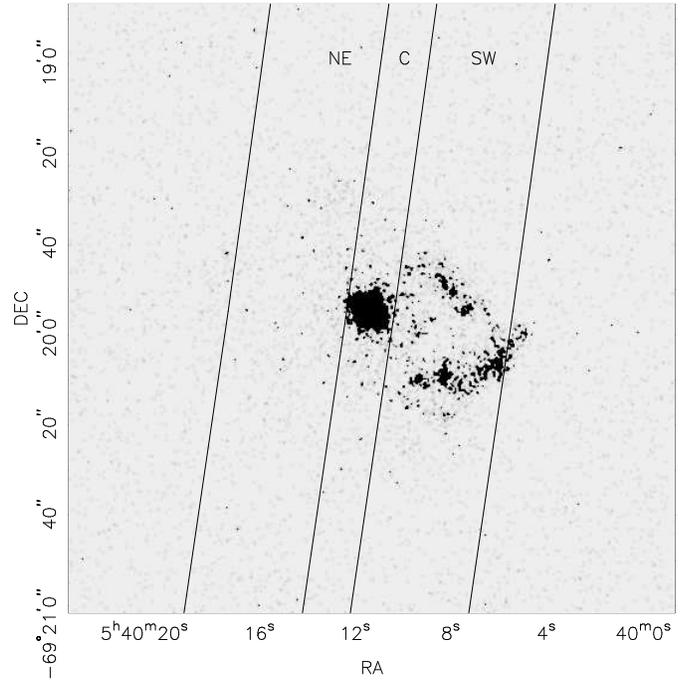}}
\caption[]{X-ray image of SNR 0540--69.3 obtained by the {\it Chandra} HRC. The image 
contains the 50 ms pulsar PSR B0540--69.3, the plerion and the outer shell. The 
SW, C and NE regions used for spectral analysis are indicated. The disperion direction 
of the RGS is along the long dimension of the extraction regions.}
\label{fig:fig2}
\end{figure}

In the SW spectrum, \ion{O}{viii} Ly$\alpha$ has a significant offset of
$\Delta\lambda = -0.17$ \AA\ with respect to the laboratory wavelength. 
This offset must be largely due to a Doppler shift, and not to a positional
offset
between the nominal pointing direction (the pulsar position) and the 
emission centroid of the shell, because the equivalent wavelength
extent of the shell is much smaller than the observed shift.
In the {\it Chandra} image, we find that the centroid of the SW 
emission, projected onto the RGS dispersion direction, is offset from
the centroid of the plerionic emission, by about $\approx -9
\arcsec$ along the dispersion direction. This corresponds to a
wavelength offset of $-0.019$ \AA. The net Doppler shift of the SW
emission is thus $-0.15$ \AA, which corresponds to a radial velocity
of $v \approx -2370$ km s$^{-1}$. A similar radial velocity is also
observed in the other strong emission lines in the spectrum. Note that
the radial velocity of the LMC, $v_{\rm LMC} = +278$ km s$^{-1}$, is
only a relatively small correction to this measurement.
The emission lines seen in the C and NE spectra are too faint to
permit a meaningful radial velocity determination.

We synthesized a spectral model comprising three components: a power law
to represent the synchrotron emission from the plerion, a non-equilibrium ionization
(NEI) model for the thermal line emission, and foreground absorption.
The free parameters of the model are the photon index of the power law and
its normalization, the electron temperature $T_{\rm e}$, emission measure, abundances,
ionization age $n_{\rm e} t$ of the shocked gas, and the column density $N_{\rm H}$
of foreground absorbing gas. Here, $n_{\rm e}$ is the electron density of the 
shell, and $t$ is the time since the hot gas was shocked.
We adopt a distance of 51 kpc throughout.

\begin{figure}
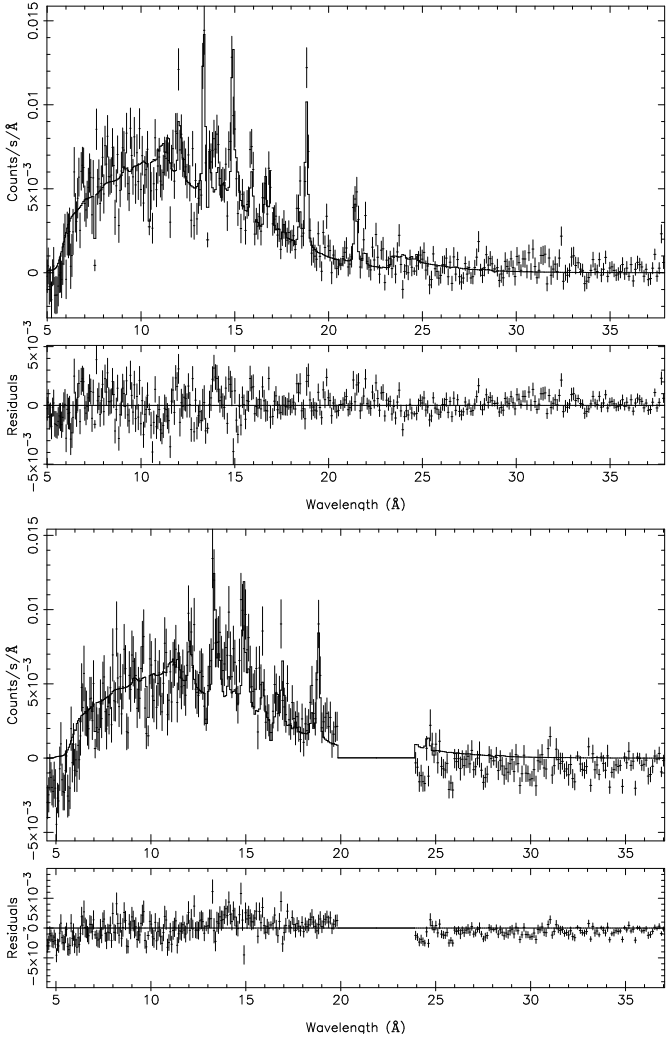

\resizebox{\hsize}{!}{\includegraphics[angle=-90]{XMM45_f3.ps}}
\resizebox{\hsize}{!}{\includegraphics[angle=-90]{XMM45_f4.ps}}
\caption[]{RGS 1 \& RGS 2 (upper and lower plots respectively) spectrum of the SW region. The solid line is a model fit to the 
data. The gap between 20 to 24 \AA\ in the RGS 2 spectra is due to the failed CCD 4 on that instrument.The spectra exhibit a strong power law continuum due to synchrotron 
radiation from the plerion and thermal line emission from highly ionized atoms 
of O, Ne and Fe }
\label{fig:fig3}
\end{figure}

We concentrated our quantitative analysis on the SW region since the spectrum from this 
region contains the strongest line emission features.  Fixing the O abundance  
to a third solar (note that we have no independent way of constraining 
the absolute abundances, since the continuum emission is dominated by
the plerionic emission), the abundances of Ne and Fe were allowed to vary with 
respect to O. An initial fit was made to the SW spectrum, using 
\ion{O}{viii} Ly$\alpha$ and \ion{O}{vii}-triplet only. The 
parameters obtained were subsequently used as starting parameters for fits to 
the entire spectrum. For the Ne/O and Fe/O abundance ratios, we obtain Ne/O $\sim 3.0$ and 
Fe/O $\sim 2.1$ relative to their solar values. The best fit model parameters 
are listed in Table 1. Fits were also made to the NE and C 
spectra.
Since the line emission in these regions is extremely weak we fixed the NEI 
component parameters (i.e. electron temperature and ionization parameter) to the 
best fit parameters obtained from the SW region, while allowing the emission 
measure and photon indices of the power law component to vary. The best fit 
parameters are given in Table 1, and the spectra, together with the 
best fit models are shown in Figures 3, 4 and 5.

As can be seen from the Figures and the $\chi^2$ values in Table 1,
the fits are not perfect. This is probably due to a combination of 
remaining small calibration uncertainties in the RGS effective area
(most of the $\chi^2$ is in the relatively poor fit to the continuum,
which is due to the plerion, at the shorter wavelengths, where the RGS
effective area is least well calibrated),
and the simplicity of our
spectral model, which does not allow for temperature and ionization 
inhomogeneities.
In particular, there is a feature at $\approx 14$ \AA, which is not fit
by our isothermal model. Candidate emission lines are 
$2p-3d$ in \ion{Fe}{xviii}
at 14.208 \AA, or possibly $2s-3p$ in \ion{Fe}{xxi} at 14.008 \AA, although the
latter should be accompanied by strong emission at 12.29 \AA\ in equilibrium.
In any case, detectable amounts of higher charge states of Fe L would definitely
indicate the presence of hotter plasma.

\begin{figure}
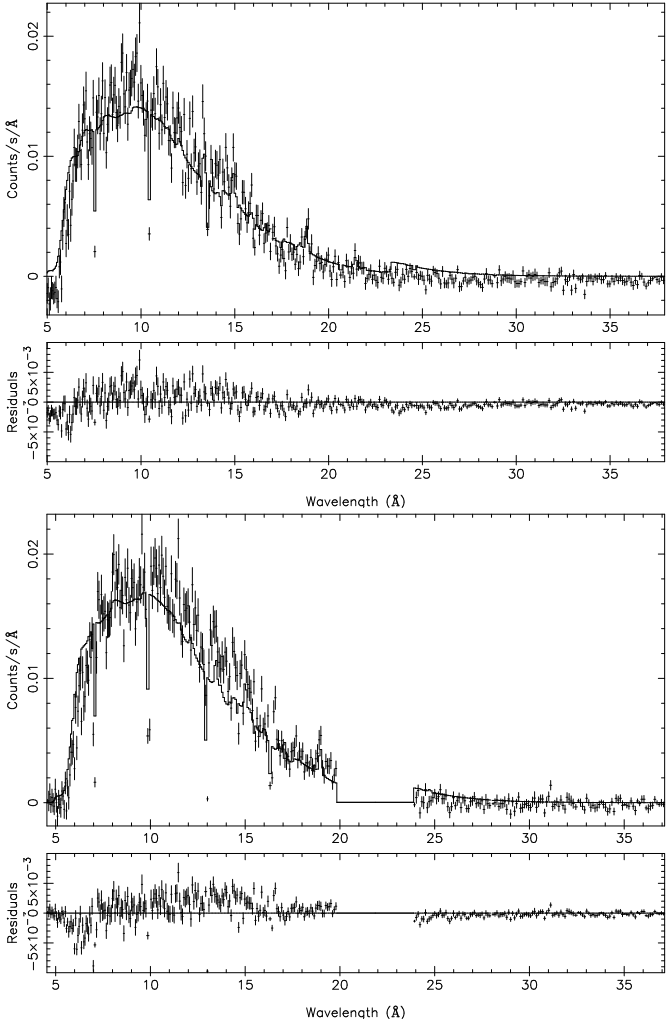

\resizebox{\hsize}{!}{\includegraphics[angle=-90]{XMM45_f5.ps}}
\resizebox{\hsize}{!}{\includegraphics[angle=-90]{XMM45_f6.ps}}
\caption[]{Same as for Figure 3, but for the C region.}
\label{fig:fig4}
\end{figure}

\begin{figure}
\resizebox{\hsize}{!}{\includegraphics[angle=-90]{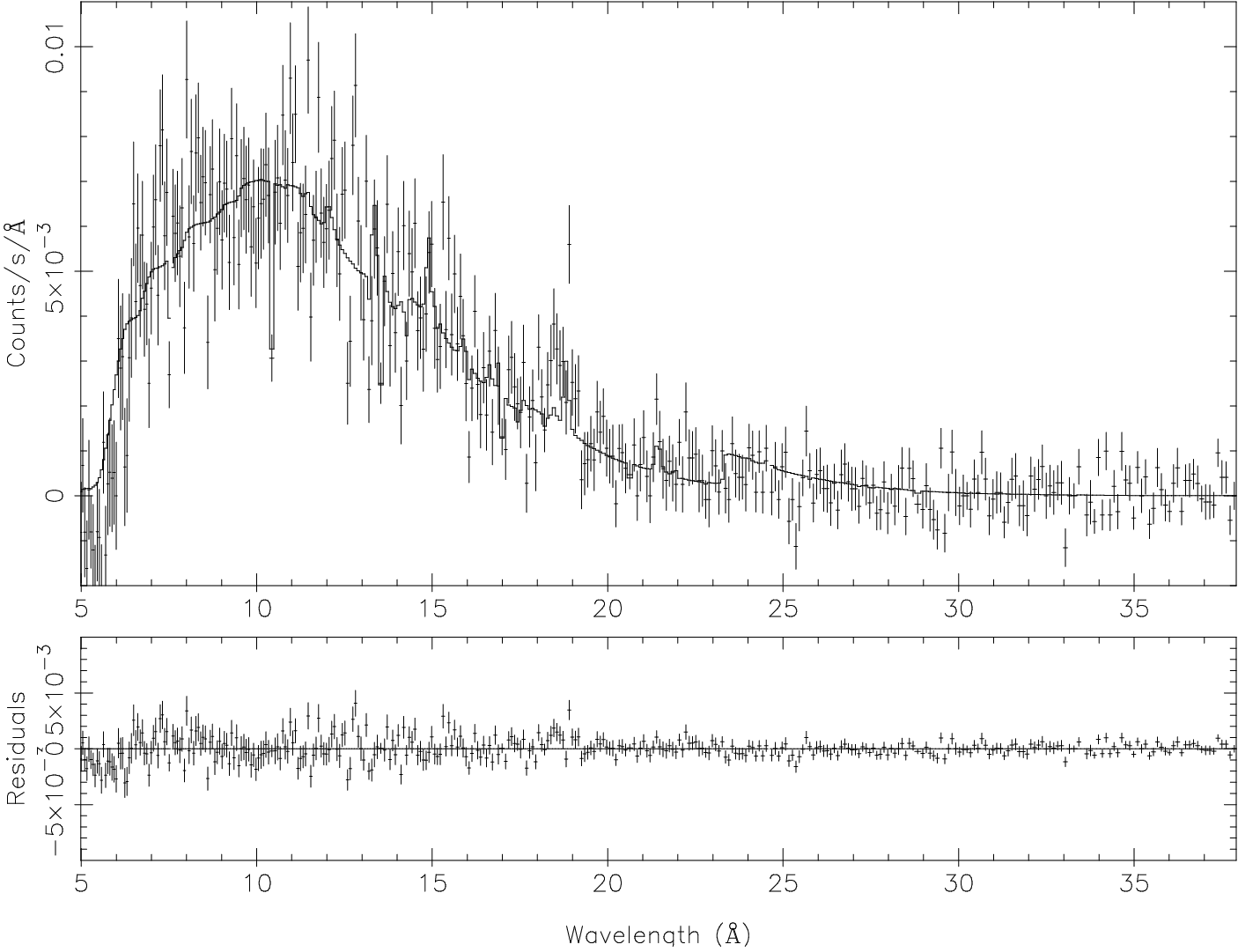}}
\resizebox{\hsize}{!}{\includegraphics[angle=-90]{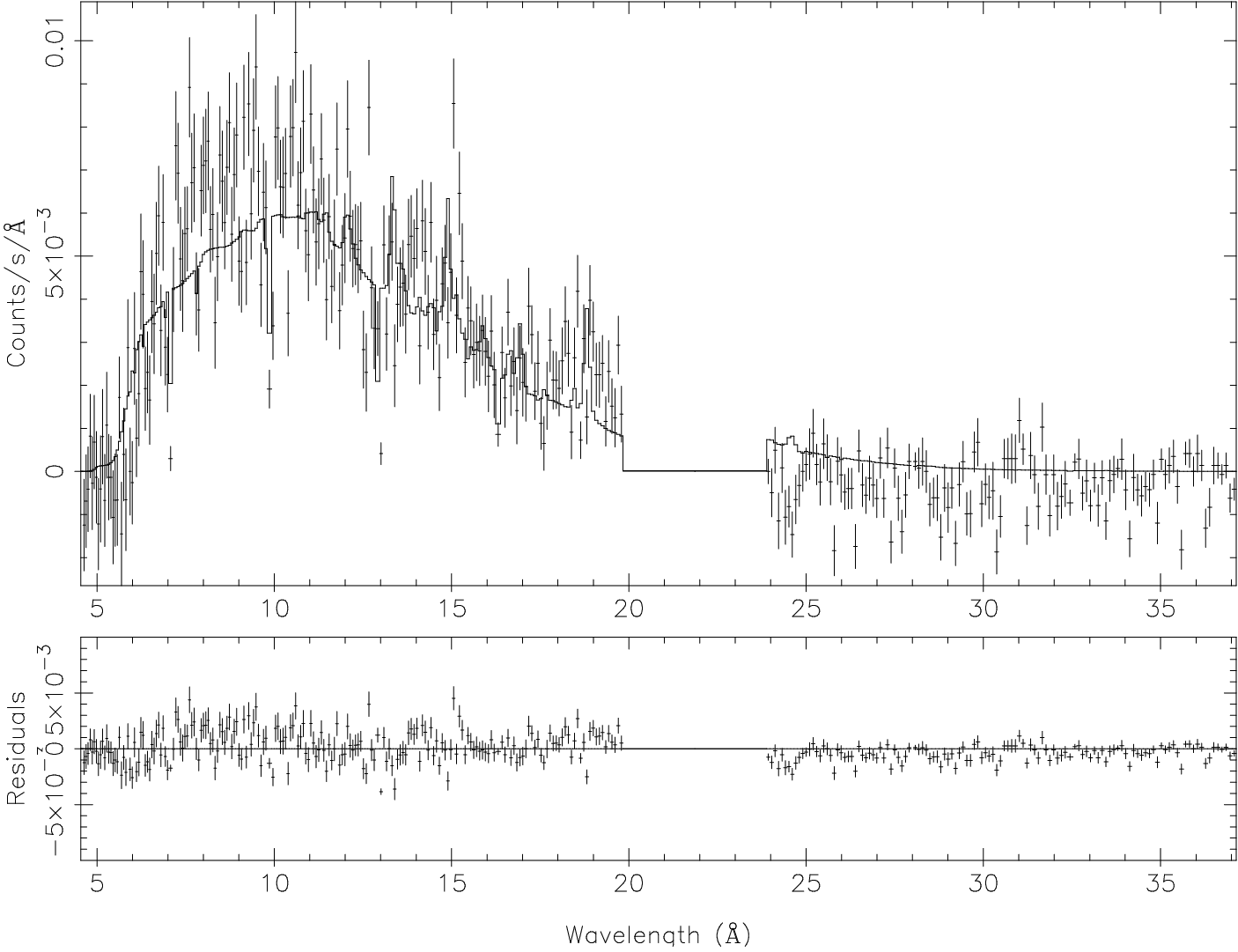}}
\caption[]{Same as for Figure 3, but for the NE region.}
\label{fig:fig5}
\end{figure}

\begin{table}[!h]
\caption{Fitting results for the SW, C and NE regions of SNR 0540--69.}
\label{tab:tab1}
\centerline{
\begin{tabular}{llll}
\hline
parameter                      &   SW         	& C & NE \\ \hline 
NEI:                           &              \\                            
$n_{\rm e}n_{\rm H}V$ (10$^{58}$cm$^{-3}$) &   92$\pm$6.5  	& 50$\pm$20 	& 45$\pm$18 \\
$kT_{\rm e}$ (keV)                   &   0.58$\pm$0.18     \\
$n_{\rm e}t$ ($10^{9}$cm$^{-3}$s)    &   25.22$\pm$5  \\
\\
POW:                           &              \\
photon index                   &   1.82$\pm$0.1 & 1.2$\pm$0.04 	& 1.75$\pm$0.1   \\
\\
ABS:                           &              \\
$N_{\rm H}$  ($10^{21}$cm$^{2}$)     &   3.88$\pm$1          \\                  
\\
\hline
$\chi^{2}$/d.o.f.              &  2.04 & 2.8 & 2.9          \\
\hline
\end{tabular}
}
\end{table}

\section{Discussion}

First of all, a qualitative point. We detect faint, but significant \ion{O}{viii}
emission from the direction of the plerion. 
At our spatial resolution, we cannot exclude the possibility that
a significant component of this line emission is in fact from
foreground remnant gas seen in projection and contamination by the SW
region. 
If confirmed at higher spatial resolution, this detection would be
interesting, because Mathewson et al. (\cite{mathewson}) detected 
intense optical [\ion{O}{iii}] emission from this same inner region, with 
a large velocity dispersion of $v \approx 3000$ km s$^{-1}$. There may
consequently be a wide 
range of ionization present in the plerion. Unfortunately,
we do not have the sensitivity to try and localize the highly ionized
oxygen, and to constrain its kinematics for comparison with the
[\ion{O}{iii}] image.

Next, we check for consistency of our measured parameter values.
The interpretation of the radial velocity measured in \ion{O}{viii}
is somewhat complicated. We see what appears to be a limb-brightened
shell in the {\it Chandra} HRC image, but if the emission is really
arranged in such a shell, one would not expect to see any radial
velocity shift of the material at all, its velocity vector being
entirely perpendicular to the line of sight. We know instead that the 
line emitting material has a velocity component along the line of
sight of $\approx -2370$ km s$^{-1}$, which, combined with an unknown
perpendicular velocity component, sets a lower limit on the
true space velocity of the \ion{O}{viii} emitting gas. To relate this
space velocity to the velocity of the expanding blast wave requires
making a further assumption about the structure and evolution of the
expanding remnant. However, the measured radial velocity still gives a
lower limit on the true expansion velocity, independent of these
assumptions.

With a measured radius of $\approx 30 \arcsec$, the shell has a linear
radius of $R_{\rm s} = 7.4$ pc. Combined with the lower limit on the
expansion velocity, we estimate an upper limit to the age of the shell
of $t < 2800$ yr, consistent with the pulsar spindown age of 
1660 yr. Conversely, 
if we assume that the age of the shell is identical to the
pulsar spindown age, we would conclude that the true expansion
velocity of the blast wave is $\approx 4600$ km s$^{-1}$, and the
radial velocity we observe in the X-ray lines is either due to
material well inside the interior of the remnant (where the flow
velocities are smaller), or the line emitting gas must have a
substantial perpendicular velocity component if it is located in the
immediate post-shock region.

The discrete emission line spectrum implies an ionization age of
approximately $n_{\rm e}t \sim 2.5 \times 10^{10}$ cm$^{-3}$ s, a low
value, which probably implies that the plasma has not yet reached 
ionization equilibrium. Combined with the pulsar spindown age, one
infers a density of the medium of $n_{\rm e} \sim 0.4$ cm$^{-3}$. But
the gas may have been shocked (much) more recently than the
characteristic age of the blast wave, depending on it position
within the remnant interior, and so we should regard this density
estimate as a lower limit. Also, from the emission measure of the
line emitting gas, we derive a density of approximately $n_{\rm e} \sim
10$ cm$^{-3}$ (assuming a half shell of relative thickness 
$\Delta R \sim R_{\rm s}/10$ for the volume estimate). Given the 
relative insensitivity of the density estimate to the volume and
emission measure, this higher density estimate may be closer to the
characteristic values at the shell than the estimate derived from the
ionization age, and this would favor the conclusion that the gas has
been shocked more recently. In fact, the measured electron
temperature, $kT_{\rm e} \sim 0.6$ keV, is incompatible with the lower
limit on the blast wave velocity $v_{\rm s} > 2400$ km s$^{-1}$
(unless the electron- and ion temperatures have not equilibrated), and
most likely, we are seeing material that has recently passed through a
(much slower) reverse shock. If gas with a much higher electron 
temperature is associated with the
outer blast wave, it will be difficult to detect it in the {\it
XMM-Newton} data. Because the remnant is only partially resolved, both
the RGS and EPIC data are dominated by strong emission from the
plerion, which reduces the sensitivity to detection of hot thermal
gas.

\section{Conclusions}

We have detected resolved
X-ray line emission from the supernova 
remnant shell around PSR B0540--69.3, confirming the nature of the
shell-like structure seen in previous experiments as a thermal SNR.
Unfortunately,
given the relatively low signal-to-noise of the observation, the fact
that the object is only partially spatially resolved, 
and the uncertainties in the interpretation of the estimates of the
various observable parameters, we can not
uniquely decide what the evolutionary status of the remnant is, or
what the relation of the line emitting gas is to the overall structure
of the remnant.

\section{Acknowledgements}
We thank Eric Gotthelf for supplying us with his {\it Chandra} HRC image. 
The Laboratory for Space Research Utrecht is supported
financially by NWO, the Netherlands Organization for Scientific
Research. 
The Columbia group is supported by the U.S. National Aeronautics and
Space Administration.
The Mullard Space Science Laboratory acknowledges financial support form the UK
Particle Physics and Astronomy Research Council.

\end{document}